# The Quartic Public Key Transformation

Gunjan Talati and Subhash Kak

**Abstract.** This paper presents the quartic public key transformation which can be used for public key applications if side information is also used. This extends an earlier work where the cubic transformation was similarly used. Such a transformation can be used in multiparty communications protocols.

**Introduction**
Cryptographic transformations may be visualized in locked box or a piggy-bank modes [1], or alternatively as one-way functions or puzzles [2],[3],[4], or as unitary transformations in quantum states [5],[6]. All these methods are variations on one-to-one mappings of different kinds. In a method that is a departure from these techniques, a three-to-one mapping for cryptography was proposed [7] with the provision that additional side-information will be provided to determine the correct message. This method was a variant of the RSA transformation where the value of the encrypting exponent was taken as 3 and $\varphi(n)$ (the totient function of the composite number $n = pq$) was divisible by 3, but not by 9. Clearly, the use of non-one-to-one mappings can make the task of the eavesdropper harder than it would be otherwise.

In the RSA transformation the decryption key is calculated by the formula $ed=1$ mod $\varphi(n)$. When Alice wants to send a message to Bob, the message to be sent is encrypted by $c=m^e$ mod $n$. Bob decrypts the ciphertext using his private key and gets the original message by $m=c^d$ mod $n=m^{ed}$ mod $n=m$. The cubic public key transformation shows how the cube roots of 1 for a prime $p$ and composite number $n = pq$ is calculated and how the three different values of messages $m$ produces the same cipher $c$. In some cases, the cube roots are obtained by applying Chinese Remainder Theorem (CRT) although congruences could also be solved by the Aryabhata Algorithm [8],[9]. Cryptography based on extraction of cubic roots from Gaussian Integers is explained in [10].

In this paper, we study quartic public key transformation as an extension of the cubic transformation. We are separately presenting additional results for general power transformations that are not one-to-one.

**The properties of the quartic transformation**
We first consider quartic mapping modulo a prime number. For quartic transformation $c = m^4$ mod $p$, four different values of message $m$ would give the same cipher $c$, where the value of prime $p$ is given by $p = 4k + 1$.

The quartic roots of 1 would be 1, $\alpha$, $\alpha^2$ and $\alpha^3$ and they are calculated by solving the equation

$$\alpha^4 - 1 = 0 \qquad (1)$$

$$(\alpha - 1)(\alpha^3 + \alpha^2 + \alpha^1 + 1) = 0$$

$$(\alpha - 1)(\alpha + 1)(\alpha^2 + 1) = 0 \quad (2)$$

The four roots obtained from the above equation are:

$$\alpha_1 = 1 \quad (3)$$

$$\alpha_2 = -1 = (p - 1) \quad (4)$$

$$\alpha_3 = \sqrt{-1} = \sqrt{p - 1} \quad (5)$$

$$\alpha_4 = -\sqrt{-1} = -\sqrt{p - 1} \quad (6)$$

Alice sends $\alpha, n$ and $c$ to Bob. When Euler totient function $\varphi(n) = p - 1$ is not divisible by 16, Bob gets to know one of the four quartic roots by the inverse exponentiation equation:

$$c^{\frac{1}{4}} = c^{\frac{a(p-1)+4}{16}} \quad (7)$$

where we pick the value of $a$ such that $\frac{a(p-1)+4}{16}$ comes out to be an integer.

Equations below show how equation (7) is derived.

$$c^{p-1} = 1 \quad \text{(Fermat's little theorem)}$$

$$c^{a(p-1)} = 1$$

$$c^{a(p-1)} c^4 = c^4 \quad \text{(multiplying } c^4 \text{ on both sides)}$$

$$c^{\frac{a(p-1)+4}{16}} = c^{\frac{1}{4}}$$

**Example 1.** Let $c = m^4 \bmod 37$. To obtain the four roots, we have to substitute the value of prime $p = 37$ in equation (4) (5) and (6).

We get the four roots as

$$\alpha_1 = 1$$

$$\alpha_2 = 36$$

$$\alpha_3 = \sqrt{37 - 1} = 6$$

$$\alpha_4 = -\sqrt{37 - 1} = -6 = 31$$

The exponent of message is 4 which is composed of only one prime 2. So, $\alpha^{\frac{4}{2}} \bmod 37 \neq 1$. Out of the four roots obtained above, 1 and 36 cannot be chosen as value of $\alpha$ because $1^2 \bmod 37 = 1$ and $36^2 \bmod 37 = 1$. We see that $\alpha$ can be either $\alpha_3$ or $\alpha_4$. Let $\alpha = 6$.

Table 1 shows the 4-to-1 mapping for prime $p=37$ and $\alpha = 6$. Table 2 explains how the communication between Alice and Bob takes place.

**Table 1**: 4 to 1 Mapping for message $m$, prime $p=37$ and $\alpha=6$

| $m$ | $m\alpha$ | $m\alpha^2$ | $m\alpha^3$ | $c = m^4 \bmod 37$ |
|---|---|---|---|---|
| 1 | 6 | 36 | 31 | 1 |
| 2 | 12 | 35 | 25 | 16 |
| 3 | 18 | 34 | 19 | 7 |
| 4 | 24 | 33 | 13 | 34 |
| 5 | 30 | 32 | 7 | 33 |
| 8 | 11 | 29 | 26 | 26 |
| 9 | 17 | 28 | 20 | 12 |
| 10 | 23 | 27 | 14 | 10 |
| 15 | 16 | 22 | 21 | 9 |

**Table 2: Communication between Alice and Bob**

| Alice | Bob |
|---|---|
| Let message $m$ chosen by Alice = 7.<br><br>After computing $m, m\alpha, m\alpha^2, m\alpha^3$ and arranging them in ascending order, Alice gets 5,7,30,32.<br><br>The rank of message $m = 7$ is 2, so the side information based on the message chosen is 2.<br><br>Alice sends $c = m^4 \bmod 37 = 7^4 \bmod 37 = 33$ and also sends side information=2 to Bob. | Using equation (7) Bob finds out that substituting a=3 gives the value of $\frac{a(p-1)+4}{16}$ as 7.<br><br>Bob gets one of the quartic roots as $33^7 \bmod 37 = 7$<br>and other roots as<br>(7*6) mod 37 = 5<br>(5*6) mod 37 = 30<br>(30*6) mod 37 = 32<br><br>Arranging them in ascending order, Bob gets 5,7,30,32 and side information=2 lets him know that the message chosen by Alice was $m = 7$. |

**Quartic transformation modulo a composite number**

Now we generalize the previous analysis to apply to composite moduli. As in RSA, the composite number, $n$ is chosen such that it is composed of two primes given by $n = pq$. We will show examples where the Euler totient function $\varphi(n) = (p-1)(q-1)$, is divisible by 4 but not by 16. The inverse exponentiation operation for quartic transformation modulo composite number is calculated by

$$c^{\frac{1}{4}} = c^{\frac{a\varphi(n)+4}{16}} \qquad (8)$$

where we pick the value of $a$ such that $\frac{a\varphi(n)+4}{16}$ comes out to be an integer.

**φ(n) divisible by 16**

When $\varphi(n)$ is divisible by 16, the sixteen quartic roots of 1 can be obtained from the equation

$$\alpha^{16} - 1 = 0 \tag{9}$$

$$(\alpha^8 + 1)(\alpha^8 - 1) = 0$$

$$(\alpha^8 + 1)(\alpha^4 + 1)(\alpha^4 - 1) = 0$$

$$(\alpha^8 + 1)(\alpha^4 + 1)(\alpha^2 + 1)(\alpha + 1)(\alpha - 1) = 0 \tag{10}$$

The message that Alice wants to send to Bob would be multiplied with sixteen roots of 1 and arranged in ascending order so that its position number in the set of 16 could be find out. After that, Alice sends $c = m^4 \bmod n$ along with the position number of message as side information.

Bob will obtain $c$. He will find the quartic root of $c$, let's say its x. Bob would then find the multiple of x with all the roots of 1, arrange them in ascending order and the position number sent as side information lets him know about the message $m$ that was sent by Alice.

**Example 2.** Let $c = m^4 \bmod 221$, $p = 17$, $q = 13$. The quartic roots for $p$ obtained by substituting $p=17$ in equation (4),(5),(6) are $\alpha_p = 1, 4, 13, 16$. The quartic roots for $q$ obtained by substituting $q=13$ similarly are $\alpha_q = 1, 5, 8, 12$

By using Chinese Remainder Theorem (CRT), we get the sixteen quartic roots of 1 as follows:

$$\alpha = (\alpha_p \times q \times ||q||_p^{-1} + \alpha_q \times p \times ||p||_q^{-1}) \bmod pq \tag{11}$$

$\alpha_1 = ((1 \times 13 \times 4) + (1 \times 17 \times 10)) \bmod 221 = 222 \bmod 221 = 1$

$\alpha_2 = ((4 \times 13 \times 4) + (1 \times 17 \times 10)) \bmod 221 = 378 \bmod 221 = 157$

$\alpha_3 = ((13 \times 13 \times 4) + (1 \times 17 \times 10)) \bmod 221 = 846 \bmod 221 = 183$

$\alpha_4 = ((16 \times 13 \times 4) + (1 \times 17 \times 10)) \bmod 221 = 1002 \bmod 221 = 118$

$\alpha_5 = ((1 \times 13 \times 4) + (5 \times 17 \times 10)) \bmod 221 = 902 \bmod 221 = 18$

$\alpha_6 = ((4 \times 13 \times 4) + (5 \times 17 \times 10)) \bmod 221 = 1058 \bmod 221 = 174$

$\alpha_7 = ((13 \times 13 \times 4) + (5 \times 17 \times 10)) \bmod 221 = 1526 \bmod 221 = 200$

$\alpha_8 = ((16 \times 13 \times 4) + (5 \times 17 \times 10)) \bmod 221 = 1682 \bmod 221 = 135$

$\alpha_9 = ((1 \times 13 \times 4) + (8 \times 17 \times 10)) \bmod 221 = 1412 \bmod 221 = 86$

$\alpha_{10} = ((4 \times 13 \times 4) + (8 \times 17 \times 10)) \bmod 221 = 1568 \bmod 221 = 21$

$\alpha_{11} = ((13 \times 13 \times 4) + (8 \times 17 \times 10)) \bmod 221 = 2036 \bmod 221 = 47$

$\alpha_{12} = ((16 \times 13 \times 4) + (8 \times 17 \times 10)) \bmod 221 = 2192 \bmod 221 = 203$

$\alpha_{13} = ( (1 \text{x} 13 \text{x} 4)+(12 \text{x} 17 \text{x} 10) ) \bmod 221 = 2092 \bmod 221 = 103$

$\alpha_{14} = ( (4 \text{x} 13 \text{x} 4)+(12 \text{x} 17 \text{x} 10) ) \bmod 221 = 2248 \bmod 221 = 38$

$\alpha_{15} = ( (13 \text{x} 13 \text{x} 4)+(12 \text{x} 17 \text{x} 10) ) \bmod 221 = 2716 \bmod 221 = 64$

$\alpha_{16} = ( (16 \text{x} 13 \text{x} 4)+(12 \text{x} 17 \text{x} 10) ) \bmod 221 = 2872 \bmod 221 = 220$

Hence, the sixteen roots of 1 are easily obtained for this example. These roots are {1, 157, 183, 118, 18, 174, 200, 135, 86, 21, 47, 203, 103, 38, 64, 220}.

Let message $m$ that Alice wants to send to Bob is 24. Multiplying 24 with sixteen roots of 1 and arranging them in ascending order, Alice obtains {10, 11, 23, 24, 28, 41, 62, 75, 146, 159, 180, 193, 197, 198, 210, 211}. The position number of 24 in the set is 4. Alice sends $c = m^4 \bmod 221$ = 55 along with side information 4 to Bob. Bob gets the cipher and calculates quartic root of 55 by using CRT which comes out to be 210. Multiplying 210 with sixteen roots of 1 and arranging them in ascending order, Bob obtains {10, 11, 23, 24, 28, 41, 62, 75, 146, 159, 180, 193, 197, 198, 210, 211}. The side information 4 lets him know that the message sent by Alice was $m$=24.

**Probability events**

The sixteen quartic roots of 1 can be put down in a different way as a number, its square and its cube.

$$1, a, a^2, a^3, b, b^2, b^3, c, c^2, c^3, d, d^2, d^3, e, e^2, e^3, f, f^2, f^3 \qquad (12)$$

Out of the nineteen values shown above 3 values would be repeated which would make 16 quartic roots of 1. They can be put into six different groups as follows:

$$1, a, a^2, a^3; \quad 1, b, b^2, b^3; \quad 1, c, c^2, c^3; \quad 1, d, d^2, d^3; \quad 1, e, e^2, e^3; \quad 1, f, f^2, f^3$$

If $x = a, b, c, d, e$ or $f$, then $x^2 \bmod n \neq 1$. If $x^2 \bmod n = 1$, then it would generate groups as $\{1, x, 1, x\}$

For the example above, we get six different groups as:

{1, 18, **103**, 86}, {1, 21, **220**, 200}, {1, 38, **118**, 64}, {1, 47, **220**, 174}, {1, 157, **118**, 183}, {1, 203, **103**, 135}

The three values which are repeated in the above sets are 103, 220 and 118.

The numbers, therefore, belong to one or more of subsets each of which has a probability of one sixth.

**Conclusions**

This article presents a method to use the quartic transformation as a public key transformation that requires the use of side information. This transformation can also be used to generate probability events that are useful in multiparty communications. Non-one-to-one mappings generalize the standard protocol for oblivious transfer which has found a host of applications and, therefore, we expect similar uses for the results of this paper.